\documentstyle[12pt]{article}
\begin{document}
\setlength{\baselineskip}{0.30in}

\newcommand{\beq}{\begin{equation}}
\newcommand{\eeq}{\end{equation}}

\newcommand{\bi}{\bibitem}


\begin{center}
\vglue .06in
{\Large \bf {Neutrinos in Cosmology}}
\\[.5in]

{\bf A.D.~Dolgov \footnote{Permanent address: ITEP, 113259, Moscow,
Russia.}, }
\\[.05in]
{\it{Institute de Fisica Corpuscular - C.S.I.C. \\
Departament de Fisica Teorica, Universitat de Valencia \\
46100 Burjassot, Valencia, SPAIN}}\\[.15in]

{Abstract}\\[-.1in]
\end{center}
\begin{quotation}
Cosmological implications of neutrinos are reviewed. The subjects considered
involve: (a) bounds on neutrino mass from the observational limits on
the universe age and the Hubble constant both in cosmology with and without
cosmological constant; (b) distortion of spectrum of cosmic neutrinos;
(c) bounds on neutrino mass from primordial nucleosynthesis; (d) lepton
asymmetry of the universe; (e) neutrino oscillations and possible new
sterile neutrinos; (f) neutrino role in large scale structure formation.
\end{quotation}

\newpage

\section{Introduction.}

There are several important subjects in cosmology where neutrinos play
(or may play) a significant role. Among them are primordial
nucleosynthesis, dark matter, and the large scale structure formation.
In the last two cases neutrinos are important only if they are massive.
Unfortunately there is no direct experimental evidence for
nonzero $m_\nu$, though the accuracy of experiments is rather loose
especially for $\nu_\mu$ and $\nu_\tau$:
\beq{
m_{\nu_e} <  5\, eV\cite{mnue}
\label{mnue}
}\eeq
\beq{
m_{\nu_\mu} <  160\, keV\cite{mnumu}
\label{mnumu}
}\eeq
\beq{
m_{\nu_\tau} <  24\, MeV\cite{mnutau}
\label{mnutau}
}\eeq
Though direct measurement give only upper bounds for $m_\nu$
there are accumulated indirect data on anomalous neutrino
behaviour which can be nicely explained by
neutrino oscillations. If this is the case, then neutrino mass should be
nonvanishing. (Inverse is not true, nonzero masses of different neutrino
flavors do not necessarily imply oscillations though the absence of
oscillations in this case would be very unnatural.)

Theory neither demands nor forbids nonzero neutrino mass. In all the
known cases of massless particles there is a theoretical principle
which not only explains the vanishing of the mass and also protects
its zero value against radiative corrections. For  example vanishing
masses of photon and graviton are ensured respectively by gauge
invariance in QCD or by coordinate covariance in general relativity.
No similar principle is known for neutrinos. So if "anything which is
not forbidden is permitted", neutrinos should be massive. One may hope
that mass spectrum of neutrinos is related to a new physics at high
energy scale beyond that of electroweak interactions. Since this scale
is unknown, theory does not say anything about possible values of
neutrino masses and one is free to speculate about them in the limits
permitted by experiment, cosmology, and astrophysics. Except for
restrictions on neutrino mass cosmology permits to put bounds on neutrino
oscillation parameters, magnetic moments, and decay properties.

\section{Relic Neutrinos and Cosmological Bounds on Their Mass.}

Neutrinos are the most abundant particles in the Universe after photons
of cosmic microwave background radiation. Their number density is
determined by thermal equilibrium which existed in the early universe
and is given by
\beq{
n_{\nu_j} = n_{\bar \nu_j} = (3 / 22) n_\gamma
\label{nnuj}
}\eeq
where $j = e,\,\mu ,\,\tau$ and
$n_\gamma =400 (T_\gamma /2.7\,K)^3 cm^{-3}$.

This result is valid if the following conditions are fulfilled:
\begin{enumerate}
\item{}Thermal equilibrium of neutrinos with the primeval electromagnetic
plasma at temperatures above 3 MeV. This is known to be true and gives
$n_\nu/ n_\gamma = 3/8$.
\item{}Adiabatic heating of photons by $e^+ e^-$-annihilation which
increases the photon number density so that the above ratio goes down
to $n_\nu /n_\gamma = 3/22$.
\item{}No other sources producing additional photons below neutrino
decoupling (at $T=3$ MeV). At sufficiently small temperatures, $T<10$ keV,
possible existence of such sources is strongly constrained by the accurate
Planck spectrum of cosmic microwave radiation.
\item{}No new interactions of neutrinos. Their frozen abundance
is determined by the usual weak interaction cross-section. If for example
neutrinos possess Yukawa interaction with massless Goldstone boson
(Majoron) with sufficiently large coupling constant $g$, their
relative number density could be as small as
$r_\nu \equiv n_\nu /n_\gamma \approx m_\nu /(g^4 m_{Pl})$ where $m_{Pl} =
1.221\cdot 10^{19}$ GeV is the Planck mass. Correspondingly the limit
obtained below for $m_\nu$ would be weaker.
\item{}Relic neutrinos are 100\% left-handed. If they have Dirac
mass and participate in
normal weak interactions only, this assumption is true roughly speaking
with the accuracy $(m_\nu /MeV)^2$. For Majorana mass no additional
states appear for any mass value.
\item{}Neutrino stability. Neutrino life-time should be bigger than the
universe age, $t_U \approx 10^{10}$ years. Otherwise their number density
at the present day would be smaller or even negligible.
Still the influence of the decay products on the universe
expansion rate permits to put some bounds on the mass/life-time even
if the decay goes into invisible particles.
\item{}Vanishing chemical potential of neutrinos. This ensures equality
$n_\nu = n_{\bar \nu}$. Normally leptonic chemical potentials are of the
same order of magnitude as the baryonic one, which is known to be very
small from the baryon asymmetry of the universe. Still strictly speaking
a large lepton asymmetry is not forbidden and in this case the limit
on the mass would be stronger. We return to the case of large neutrino
chemical potentials in connection with primordial nucleosynthesis.
\end{enumerate}

Energy density in the universe is characterized by the cosmological
parameter $\Omega = \rho /\rho_c$ where $\rho_c = 3H^2 m_{Pl}^2 /8\pi$
and $H$ is the Hubble constant which is parametrized as
$H=100h_{100}\,km/sec/mpc $. Since the energy density of neutrinos
(and antineutrinos) is smaller than the total energy density of
matter we can
write the upper bound on their mass (Gerstein-Zeldovich limit)\cite{gz}
as:
\beq{
\sum_j m_{\nu_j} < \Omega_m \rho_c /(n_\nu + n_{\bar \nu}) =
94 h_{100}^2 \Omega_m \,eV
\label{mnugz}
}\eeq
Here $\Omega_m$ corresponds to energy density of matter in contrast to
vacuum energy density (or cosmological constant). Since $\Omega$ is not
a well known quantity it could be more restrictive to express the bound
on $m_\nu$ through the lower limit on the universe age and the Hubble
parameter. The universe age is given by
\beq{
t_U ={1 \over H} \int^1_0 {dx \over (1-\Omega_{tot} +
\Omega_m /x + \Omega_r/ x^2  + \Omega_v x^2)^{1/2} }
\label{tu}
}\eeq
where $\Omega_m$, $\Omega_r$, and $\Omega_v$ are the present-day
fractions of cosmological
energy density of nonrelativistic matter, relativistic matter, and vacuum
(cosmological constant); $\Omega_{tot} = \Omega_m + \Omega_r + \Omega_v$.
In spatially flat universe, as advocated by inflationary scenario,
$\Omega_{tot}=1$. It is usually assumed that $\Omega_r \ll \Omega_m$
because relativistic energy density decreases faster in the course of
expansion than the nonrelativistic one. It may be not so if there are
late-decaying particle producing relativistic decay products at
contemporary epoch. It is also assumed that $\Omega_v=0$ (vanishing
cosmological constant). This assumption has no theoretical justification,
moreover any reasonable theoretical estimate gives the value of vacuum
energy density 50-100 orders of magnitude bigger than the observational
limit\cite{sw}. So having something so small, because of unknown mechanism,
one assumes that this quantity is exactly zero. The recent conflict between
a large universe age and a high value of the Hubble constant indicates
that cosmological constant and correspondingly $\Omega_v$ might be nonzero.
The universe age is determined from the ages of old globular clusters and
the relative abundances of long-lived radionucleides and is found to be
in the range $t_U= 12-15$ Gyr. Recently even a larger value, $t_U=17$ Gyr,
was advocated (for the review see \cite{vdb}). The Hubble parameter is
probably somewhere between $0.5<h_{100}<1$. The new data has tendency to
higher values, $h_{100}= 0.7-0.8$\cite{hubble} but is still hard to
estimate systematic errors.

Assuming that good old cosmology with zero cosmological term is valid and
approximating the integral (\ref{tu}) by the expression
$t_U\approx [H(1+\sqrt{\Omega}/2)]^{-1}$ we get
\beq{
\sum_j m_{\nu_j} < 390\,eV \left( {9.8\,Gyr \over t_U } -h_{100} \right)^2
\label{mnutu}
}\eeq
With $h_{100}=0.65$ and $t_U > 12$ Gyr one gets $m_\nu < 10$ eV. With larger
$H$ and $t_U$ the bound is even stronger but at some stage the assumption
of vanishing $\Omega_v$ becomes incompatible with their high values and
one has to invoke nonzero cosmological constant. The bound becomes weaker
but still meaningful. For example with $\Omega_{tot}=1$, $t_U > 14$ Gyr,
and $h_{100} > 0.75$ we get $m_\nu < 20$ eV.

There is also the well known bound on the mass of a very heavy neutrino
(if it exists) from below\cite{dvz}:
$m_\nu > 3$ GeV. It was obtained with $\Omega_v =0$. Relaxing this assumption
one gets the limit 2-3 times weaker. These limits are not very interesting
after direct measurement of the decay width of $Z^0$ made at LEP which
showed that there is no space for an extra neutrino with mass below $m_Z/2$.
Moreover it is difficult to believe that so heavy neutrinos could be stable
on cosmological time scale, though formally it is not excluded.

\section{Spectrum of Cosmic Neutrinos}

It is assumed usually that cosmic neutrinos (if they are massless)
have equilibrium Fermi-Dirac spectrum with vanishing chemical
potentials:
\beq{
f_\nu = 1/[\exp(E/T_\nu) + 1]
\label{specnu}
}\eeq
with the temperature $T_\nu = (4/11)^{1/3} T_\gamma = 1.93(T\gamma/2.7)$.
However in contrast to electromagnetic background radiation where spectral
distortion is extremely small, below $10^{-4}$\cite{cobe}, neutrino spectrum
is much more distorted. It is because electrons and neutrinos have different
temperatures at $T<m_e$ so that the annihilation $e^+ e^- \rightarrow
\bar \nu \nu$ produces nonequilibrium $\nu$ and $\bar \nu$ which cannot
thermalize at these low temperatures. Calculations of ref.\cite{df} give
the result:
\beq{
\delta f_{\nu_e} /f_{\nu_e} \approx 5\times 10^{-4} (E/T) (11E/4T -3)
\label{deltaf}
}\eeq
Numerical calculations\cite{dt} give similar results. The distortion
for $\nu_\mu$ and $\nu_\tau$ is approximately twice smaller because
at that temperatures they have only neutral current interactions.

This effect results in an increase of neutrino number density at
the present day by almost 1\%. It is not important from the point of
view of the bound on their mass. It could be potentially essential for
the primordial nucleosynthesis. Distortion of the electronic
neutrino spectrum would change the neutron-to-proton ratio
because electronic neutrinos (in contrast to $\nu_\mu$ and $\nu_\tau$)
participate in the reactions $n+\nu \leftrightarrow p+e^-$ and
$p+\bar \nu \leftrightarrow n+e^+$ and directly shift its value
(not only through the influence on the cooling rate). If there is an
excess of $\nu_e$ and equally of $\bar \nu_e$ at higher energies
(with respect to the equilibrium values) the $n/p$-ratio would be
bigger because the number density of protons is larger than the number
density of neutrons by factor $\exp (\Delta m /T)$ and correspondingly
destruction of neutrons in the first reaction is
less efficient than the production of them in the second reaction. An
excess of neutrinos at low energy produces the opposite effect because
of threshold 1.8 MeV in the second reaction which inhibits neutron
production. The correction (\ref{deltaf}) could shift the $n/p$-ratio
at per cent level but for this particular case
its influence on nucleosynthesis is practically negligible. As we have
mentioned above the dependence of $n/p$-ratio on the spectrum
corrections is not sign-definite and it
happened that the spectrum was distorted in such a way
that $n/p$-ratio does not change. The effect would be much
bigger if nonequilibrium $\nu_e$ ($\bar \nu_e$)
come from the annihilation of heavy tau-neutrinos with the mass around
10 MeV, $\nu_\tau \bar \nu_\tau \rightarrow \nu_e \bar \nu_e$\cite{dpv}.

Another possible
source of nonequilibrium electronic neutrinos could be decays
of massive particles\cite{ts,dk} after neutrino decoupling from
the cosmic plasma, that takes place around 2 MeV. A possible candidate for
the role of the mother-particle is massive $\nu_\tau$. As we
mentioned above the effect of nonequilibrium $\nu_e$ could shift
$n/p$-ratio either way. In particular if $n/p$ goes down, this would
relax the Schwartsman bound\cite{schw} on the number of massive
neutrino species\cite{dk} or relax the bound on
the baryon-to-photon ratio during nucleosynthesis\cite{gt}.

\section{Bounds on Neutrino Mass from Nucleosynthesis}

In the case that neutrinos live longer than nucleosynthesis time,
$t_{NS}\sim 100$ sec but shorter than the universe age,
$t_U \sim 10^{10}$ years, consideration of primordial nucleosynthesis
permits to exclude an interesting interval of $\nu_\tau$ mass\cite{ckst,
dr}, while for $\nu_e$ and $\nu_\mu$ the bounds are weaker than the
experimental ones (1,2). The arguments are essentially the same as those
leading to the nucleosynthesis bound on the number of massless neutrino
flavors\cite{schw}.
New particle species in the primeval plasma during nucleosynthesis
epoch would change the universe cooling rate and correspondingly the
frozen value of neutron-to-proton ratio which predominantly
determines abundances of the produced light elements. Concordance with
observations leads to the bound on the extra neutrino species,
$k_\nu < 1$. Quite recently the bound was more restrictive, $k_\nu<0.3$
or even $k_\nu <0.1$ but recent data on primordial $^4He$ and deuterium
created some confusion and the relaxation of the bound. For the details
and references see the talk by G. Steigman at this conference\cite{gs}.

If neutrinos are heavy their influence on the cooling rate would be
similar to addition of extra massless neutrino flavors.
Though in equilibrium the energy density of massive particles
is smaller than that of massless ones, tau-neutrinos with mass in
MeV-range went out of equilibrium when their number density is still
nonnegligible and since the energy density of nonrelativistic particles
in the course of expansion decreases more slowly, they gradually begin
to dominate. For example 10 MeV tau neutrino which is stable on the
nucleosynthesis time scale is equivalent to almost 7 massless neutrino
species if it is a Dirac particles and to 4 species if it is a Majorana
one \cite{dr}. This argument permits to exclude $\nu_\tau$ in the mass
interval $0.5< m_{\nu_\tau}<35$ MeV\cite{ckst,dr} if the permitted bound
on extra neutrino species is $\Delta N_\nu < 0.6$. In the case of a
weaker bound, $\delta N_\nu <1$, the excluded mass interval shrinks to
$1<m_{\nu_\tau}<30$ MeV. These results are valid for the Majorana type
neutrinos. In the Dirac case the lower limits are approximately twice
better. It is connected with twice larger number of possible states for
the Dirac particles. The occupation number of right-handed neutrinos
in the primeval plasma was calculated in ref.\cite{dkr} (for earlier
papers see\cite{mnud}) where it was shown that in the case of a very
strong bound $\Delta N_\nu < 0.1$ the lower end of the excluded interval
for the Dirac tau neutrino goes down to approximately 10 KeV.

These results were obtained under assumptions of kinetic equilibrium
of neutrinos in the primeval plasma. As is mentioned in the previous
section this assumption is violated and nonequilibrium electronic
neutrinos may considerably strengthen the limits. For example
nonequilibrium $\nu_e(\bar \nu_e)$ coming from the annihilation of 20
MeV tau neutrinos are equivalent to almost one extra neutrino species
if $\nu_\tau$ has the Dirac mass and to 0.15 extra nus if it has the
Majorana mass\cite{dpv}.

Tau-neutrinos with MeV mass would not spoil successful nucleosynthesis
results if they are unstable on the
nucleosynthesis time scale. This case was
analyzed in ref.\cite{unstable}. The bounds on the mass depend upon
the life-time and decay channels. For a sufficiently short life-time
tau-neutrinos remain in equilibrium during nucleosynthesis, their number
density is Boltsmann suppressed and practically any mass is permitted.
Another way to avoid the nucleosynthesis bound on the mass is to assume
a new interactions for $\nu_\tau$ which could deplete their density at
nucleosynthesis. Since MeV tau-neutrinos should be unstable anyhow a new
interaction which generates the decay, is necessary. This could be flavor
nonconserving effective four-fermion interaction generating the decay
$\nu_\tau \rightarrow 3\nu$ or the Yukawa coupling to a light (or massless)
scalar boson (like e.g. Majoron) producing the decay
$\nu_\tau \rightarrow J+\nu_l$ where $l$ stands for $e$ or $\mu$. The
life-time with respect to these decays may be very long so that $\nu_\tau$
remains stable during nucleosynthesis. It is possible that the nondiagonal
coupling $g'\nu_\tau \nu_l J$ leading to the decay is much weaker than
the diagonal one $g \nu_\tau \nu_\tau J$. In this case the annihilation
$\nu_\tau+\nu_\tau \rightarrow 2J$ may be efficient during nucleosynthesis
diminishing $\nu_\tau$ number density\cite{dprv}.
This could help to avoid the mass limits mentioned above. For the details
and references see also the talk by S.Pastor at this conference.

\section{Lepton Asymmetry}

It is usually assumed that there is no charge asymmetry in lepton sector,
the number density of neutrinos is equal to that of antineutrinos. Lepton
asymmetry is not directly observable and in principle may be large. The
reason for the assumption of its smallness is a small value of the baryon
asymmetry, $(n_B - n_{\bar B}) /n_\gamma \approx 3\times 10^{-10}$. Usually
theoretical models predict lepton and baryon asymmetry of about the same
magnitude though there may be interesting exceptions.
The value of charge asymmetry in kinetic
equilibrium can be characterized by chemical potential $\mu$ so that
the expression (\ref{specnu}) is changed to
\beq{
f_\nu = 1/[\exp((E-\mu)/T_\nu) + 1]
\label{mns}
}\eeq
In chemical equilibrium $\bar \mu + \mu =0$ where $\bar \mu$ is
the chemical potential of antiparticles.
It is convenient to introduce the quantity $\xi=\mu/T$ which remains
constant in the course of expansion if the corresponding charge is
conserved.

The strongest bound on the magnitude of lepton asymmetry can be derived
from primordial nucleosynthesis. Nonzero chemical potential results in
an increase of neutrino energy density and in this sense is equivalent
to an addition of extra neutrino species. If the nucleosynthesis upper
bound is $\Delta n_\nu <1$ then $|\xi_l| < 1.5$, and if $\Delta n_\nu <0.3$
then $|\xi_l| < 0.8$. For electronic type neutrinos the limit is
much stronger because, as we have mentioned above, the $n/p$-ratio
is especially sensitive to the spectrum of electronic neutrinos. If the
data on light element abundances permit one extra neutrino species
electronic chemical potential is bounded by $\xi_e < 0.07$. For more
details and the list of references see review paper\cite{ad3}.

Lepton asymmetry should be generated along the same lines as the
baryon asymmetry, namely by the out-of-equilibrium processes with
leptonic charge nonconservation and C and CP breaking\cite{ads}.
Leptogenesis in GUT models predicts lepton symmetry of the same
order as the baryonic one and correspondingly $\xi \leq 10^{-9}$.
Electroweak leptogeneration\cite{krs} satisfies the condition of
$(B-L)$-conservation and also predicts a very low result for the
lepton asymmetry. Moreover if electroweak phase transition is second
order then the asymmetry is not generated but washed out by
electroweak processes. In this case any preexisting state with
arbitrary $B=L$ acquires $B=L=0$ after electroweak stage. However
if there was a primordial lepton asymmetry $L_i$ then after
electroweak epoch asymmetry $B=L=L_i/2$ would be generated.
The initial lepton asymmetry might come from the
out-of-equilibrium decays of heavy Majorana neutrino\cite{fy}.
One sees that in this case the lepton asymmetry is very small too.
Still there is a hope to generate a large lepton asymmetry in a
version\cite{dk2} of the model of baryogenesis with baryonic (and
leptonic) charge condensate\cite{afd} (see also \cite{ad3}). It is
essential that electroweak processes would not spoil this result;
this could happen if the relevant processes take place below the
electroweak scale or if electroweak baryo- and lepto- genesis do not
operate. The model\cite{dk2} predicts
a relatively small spatial scale of the variation of lepton
asymmetry. The concrete size of the scale is model dependent and
could quite easily be as small as O(kpc) or as large as O(Gpc). It
is interesting if the recently observed\cite{deut} different abundances
of primordial deuterium at large distances,
$z\approx 3$, could be explained
by variation of leptonic chemical potentials. If this is the case
then not only deuterium but other light elements (especially $^4 He$)
should have systematically varying abundances.

\section{Neutrino Oscillations and New Neutrinos}

Neutrino oscillations is probably the central (hypothetical) phenomenon
in neutrino physics. There is not yet conclusive laboratory evidence in
favor of oscillations but an impressive experimental activity in the
field makes one hope for an essential progress in the near future. For
the reviews and references see talks by Caldwell\cite{cald} and
Valle\cite{valle} at this conference. There are plenty of indirect
evidence in favor of oscillations. First among them is the deficit of
solar neutrinos discovered by the group led by Davies whose 80th
anniversary we all celebrate here. This deficit may be explained by
the resonance neutrino oscillations (the MSW effect) and there is
plenty of discussion of the problem at this conference. There are some
other observed anomalies in neutrino physics like atmospheric neutrino
problem or Karmen anomaly. If all the data are correct the implications
could be quite exciting. One possibility is an existence of a new
sterile neutrino, $\nu_s$  with an efficient oscillations between
$\nu_s$ and normal neutrinos (for a recent discussion see\cite{pelt}).

Oscillations into new neutrino states would distort successful
nucleosynthesis predictions and hence a bound on oscillations parameters
can be derived. Neutrino oscillations in the hot dense cosmic plasma
at high temperatures when neutrinos are strongly coupled to the plasma
is a rather comlicated phenomenon. It cannot be described by the usual
Schroedinger equation but the density matrix formalism
should be used instead\cite{ad4}. At smaller temperatures (roughly
speaking below 2-3 MeV), when neutrino scattering dies down, one can
return to the Schroedinger equation with the properties of the medium
described by refraction index\cite{raf}. The bounds on the oscillation
parameters derived in refs.\cite{bd} are meaningful if primordial
nucleosynthesis strongly constraints the number of extra neutrino
species. In the case that $\Delta n_\nu =1$ is permitted no restriction
follows from nucleosynthesis for oscillations into one and only one new
state. It was argued in ref.\cite{fv1} that neutrino oscillations could
give rise to a large leptonic chemical potentials. This is a very interesting
result though it is in contradiction with papers\cite{bd}. It deserves 
further consideration.

Recently there appeared a renewed interest\cite{bm,fv,sil,bdr} to the old idea
of the mirror world\cite{miro}. It is assumed that there exists another
world almost or exactly symmetrical to ours which is coupled to our world
only through gravity and possibly through a new very weak interaction.
Such a possibility is inspired by superstrings with the symmetry group
$G_{tot}=G\times G'$ (like e.g. $E_8\times E'_8$). Another world contains
the same set of particles and similar interactions. Exact symmetry
between the two worlds are forbidden by the nucleosynthesis because the\
mirror world contributes effectively $\Delta n_\nu = 10.75/1.75 = 6.14$
to the number of massless neutrino species. However if the symmetry is
broken so that the (re)heating temperatures after inflation are different
the nucleosynthesis constraints may be satisfied. A very interesting
phenomenologically model arises if the electroweak symmetry  breaking
scales are different in our and mirror worlds, $v'/v \approx 30$. To
satisfy the nucleosynthesis constraints in this case the ratio of the
temperatures of normal and mirror particles at the nucleosynthesis era
should be $T'/T < 0.96 (\Delta n_\nu)^{1/4}$\cite{bdr}. Oscillations
between mirror and normal neutrinos with
a reasonable choice of parameters may explain all known neutrino
anomalies\cite{bm}. The model predicts the existence of relatively
heavy mirror neutrinos with the mass in keV range which may be warm
dark matter while light neutrinos with mass in eV range are natural
candidates for hot dark matter. Because of different electroweak
scales the masses of fundamental fermions in the mirror world are
about 30 times larger. This results in the absence of stable mirror
nuclei and in turn to quite different astrophysics and in particular
to an easier black hole formation.

If neutrinos have a magnetic moment then their interaction with magnetic
field would result in spin-flip and so normal left-handed neutrinos
would be transformed into sterile right-handed ones. If this process was
efficient during nucleosynthesis it doubled the number
of neutrino species. Assuming that during nucleosynthesis there
existed magnetic fields in the primeval plasma which seeded the
present-day magnetic fields in galaxies, one can put an upper bound on
the neutrino magnetic moment, $\mu_\nu <10^{-16} \mu_B$\cite{ers} where
$\mu_B$ is the electron Bohr magneton.

\section{Neutrinos and the large Scale Universe Structure}

Massive neutrino are natural candidates for dark matter particles. In
comparison with other candidates neutrinos have a definite advantage:
they are known to exist and it is natural to expect that they are
massive. Unfortunately the theory of large scale structure formation with
light neutrinos ($m=O(10)$eV), so called hot dark matter, does not fit
the observed picture. Formation of galactic (and smaller) size structures
is strongly suppressed. Moreover light neutrinos contradict Tremaine-Gunn
limit\cite{tg}. Because of Fermi exclusion principle one cannot squeeze
arbitrary many neutrinos into a galaxy and to represent invisible mass
in dwarf spheroids they should be heavier than 0.3-0.5 keV. So
one has to invoke new heavy hypothetical particles (cold dark matter).
Cosmology presents one of the strongest arguments in favor of their
existence and thus of new physics beyond the standard model. (For
the recent review see e.g. ref\cite{ad5}). Still neutrinos are probably
not absolutely useless for structure formation. One of the popular models
requests 70\% of cold dark matter and 30\% of hot dark matter
and might be especially good if there are two equal mass neutrinos with
$m=2.5$ eV\cite{cald}.

A single  component dark matter model looks of course more natural and
attractive. Unfortunately with the simple assumption of scale-free (flat)
spectrum of initial perturbations\cite{hz} such  models do not agree
with observational data. Addition of 30\% of hot dark matter permits to
increase the power at large scales without distortion small scales.
The same goal can be achieved with heavy (MeV) unstable but long lived
particle\cite{unst}. A rather natural candidate for such a particle
is tau-neutrino. The role of the decaying particle is to enhance the
energy density of relativistic matter coming from the product of its
decay. This would result in a later onset of nonrelativistic stage
and correspondingly to a smaller power of evolved structures at small
scales. An improvement of the bound on $\nu_\tau$ mass is very interesting
from the point of view of testing these models.

\section{Acknowledgement}
This work was supported by DGICYT under grants PB92-0084 and SAB94-0089.

\end{document}